\documentclass[a4paper]{spie}

\usepackage{amsmath,amsfonts,amssymb}
\usepackage{graphicx}
\usepackage{siunitx}
\usepackage{array}
\usepackage{textcomp}
\usepackage{subcaption}
\usepackage{float}

\title{4D blood flow mapping using SPIM-microPIV in the developing zebrafish heart}

\author[a]{Vytautas Zickus}
\author[a]{Jonathan M. Taylor}
\affil[a]{University of Glasgow, School of Physics and Astronomy, Glasgow, G12 8QQ, UK}

\authorinfo{E-mail: Jonathan.Taylor@glasgow.ac.uk}
\begin{document}
\maketitle

\begin{abstract}
Fluid-structure interaction in the developing heart is an active area of research in developmental biology. However, investigation of heart dynamics is mostly limited to computational fluid dynamics simulations using heart wall structure information only, or single plane blood flow information - so there is a need for 3D + time resolved data to fully understand cardiac function. We present an imaging platform combining selective plane illumination microscopy (SPIM) with micro particle image velocimetry ({\textmu}PIV) to enable 3D-resolved flow mapping in a microscopic environment, free from many of the sources of error and bias present in traditional epifluorescence-based {\textmu}PIV systems. By using our new system in conjunction with optical heart beat synchronisation, we demonstrte the ability obtain non-invasive 3D + time resolved blood flow measurements in the heart of a living zebrafish embryo.
\end{abstract}

\keywords{Selective Plane Illumination Microscopy,  Micro Particle Image Velocimetry,  Correlation Averaging, Depth of Correlation,  Blood Flow Imaging,  Zebrafish, Optical Gating}

\section{INTRODUCTION}
\label{sec:intro}  
Micro Particle Image Velocimetry ({\textmu}PIV) is a non-invasive flow measurement technique, which uses image cross-correlation to quantify the motion of groups of particles in a fluid\cite{adwest_piv_2011, Willert2007}. {\textmu}PIV has been utilized increasingly in recent years to quantify micro-scale flows in a range of different applications \cite{Lindken2009,Wereley2010}, including flow related to biological systems such as the swimming zooplankton \cite{Gemmell2014}, root growth \cite{Bengough2009}, blood sucking mosquito \cite{Kikuchi2011} and, most relevant for this work, measurement of blood flow in small animal models such as the rat\cite{sugii_2002}, chicken\cite{Vennemann2006,poelma_shear}, and zebrafish\cite{Hove2003,Forouhar2006,Ohn2009,Lu2010,Jamison2012}.

Previous quantification of red blood cell (RBC) dynamics in such animals using {\textmu}PIV measurements has mostly been limited to two-dimensional two-component (2D-2C) flow information obtained from epifluorescence or brightfield (BF) microscopy. However, not only does this fail to capture the full 3C-3D flow field, these volume illumination imaging modalities also suffer from velocity underestimation when investigating fluids with a depth-varying velocity profile \cite{Kloosterman2011}, such as the blood flow in the heart.
Conceptually the reason for this is the presence of an extended depth over which particles outside of the depth of focus (DoF) still strongly influence and bias the measured flow values. This can be interpreted as a weighted averaging of the true flow velocity value throughout the thickness of the sample. This depth depends on the imaging system and the particles in the fluid\cite{Olsen2000}, and even on the dynamics of the imaged flow itself \cite{Olsen2003}, and is referred to as the Depth-of-Correlation (DoC).

Due to these limitations of DOC and lack of optical sectioning to resolve in depth, standard {\textmu}PIV using epifluorescence microscopy only provides a ballpark estimate of the true underlying flow. To overcome these issues, and also to extend to direct 3D-3C measurements, several techniques have been proposed (reviewed in detail by Cierpka \& Kehler\cite{Cierpka2012a}).  All of these systems have their own limitations, including, but not limited to, the need for multiple imaging arms, high photo-toxicity to living samples, only applicable for certain densities of flowing particles, lowering of SNR, and/or the need for non-trivial algorithms for flow measurement reconstruction.

In this manuscript, we will describe how we have used a selective plane illumination microscopy (SPIM) system to overcome these limitations and acquire 3D-2C flow fields using a method that is highly compatible with living samples. We demonstrate phantom experiment results, which illustrate the robustness of the system and ability to obtain 3D-2C data directly. We experimentally compare flow measurements in a circular tube with BF and SPIM, where DoC effects are clearly manifest in the recovered flow profiles in the BF case, but are absent for SPIM. We also quantify the effect of out-of-plane motion (OOPM) effects, and discuss how correlation averaging can help minimize them by improving the statistical accuracy of the measurements.
Finally, we will demonstrate how the gentle imaging of SPIM can be combined with our previously-reported optical gating techniques \cite{Taylor2012} to obtain high quality time- and depth-resolved blood flow measurements in the heart of a zebrafish embryo. Our 3D-2C measurements are robust against noise, low blood cell density, and depth-varying flow.

\section{Method}
\begin{figure}[b]
	\centering
	\includegraphics[width =\textwidth]{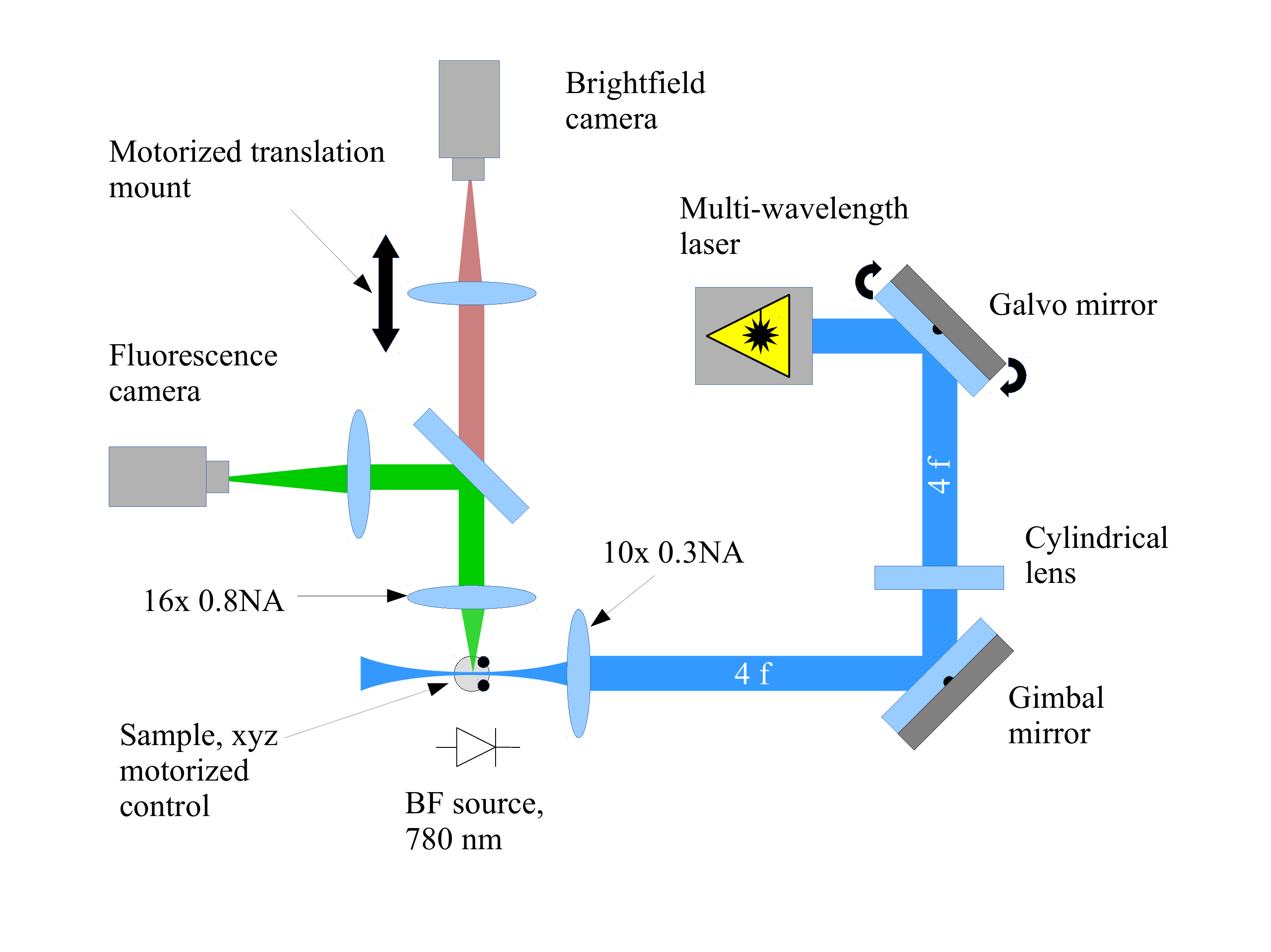}
    \caption{A simplified diagram of the SPIM system used in this work, a more detailed diagram can be found in \cite{Taylor2012}. Like in macroscopic PIV systems, the light sheet is formed using a cylindrical lens, which allows for instantaneous large FOV illumination. Simultaneous imaging of the zebrafish heart with brightfield (BF) and fluorescence (FLR) channels allows for the use of optical gating techniques \cite{Taylor2014, Liebling2005}. The 780 $\si\nm$ LED serves as the BF illumination source, and the BF channel uses a motorized tube lens to keep the image in focus, as the sample is moved through the light sheet. A galvo mirror is used to minimize shadow artefacts \cite{Huisken2007}. The 4f relay optics are omitted from the diagram for clarity.}
     \label{fig:spim_diagram}
\end{figure}
\label{sec:method}
The volume illumination effects in {\textmu}PIV measurements mentioned in Sec.~\ref{sec:intro} are an issue for the epifluorescence and BF microscopy modalities that have been used in the past to image blood flow in the zebrafish heart.
The SPIM system used in this work, Fig.~\ref{fig:spim_diagram}, has optical similarities with conventional macroscopic PIV systems, in that both use a cylindrical lens to create thin sheet of light to illuminate the sample. Because the illumination is confined to the thickness of the light sheet, these DoC effects do not occur, and true depth sectioning is achieved.

All PIV analysis was performed using the OpenPIV software \cite{openpiv2010}. Frame pairs were acquired from the fluorescence imaging channel, closely-spaced in time, and processed using a standard PIV analysis. This involves dividing up the images into small ``interrogation windows'', within which the two images in a pair are cross-correlated to infer a local motion vector. We used sum of absolute difference (SAD) of pixel intensity values as our correlation function (as opposed to conventional sum of pixel intensity product) as we found empirically that it performs better for RBC imaging.
We note that there is a vast array of pre- and post processing methods in PIV literature but, unless otherwise stated, we have not used any such techniques in order to illustrate the fact that the quality of the raw images alone is sufficient to obtain smooth flow field information.
\begin{figure}[b]
\centering
\subcaptionbox{\label{bf-tube}}{\includegraphics[trim = 5mm 5mm 5mm 5mm, width=.49\textwidth]{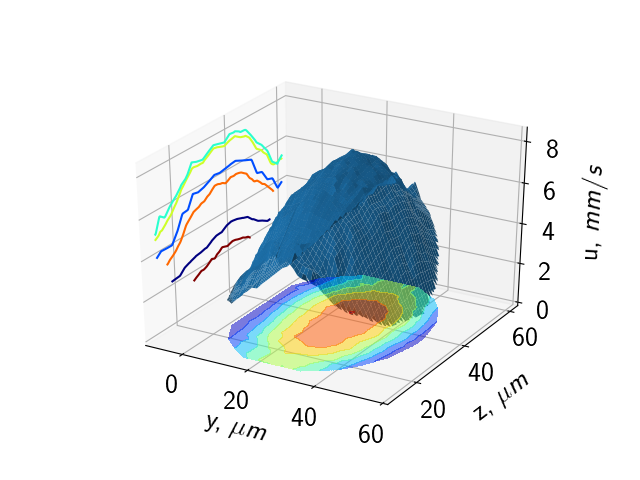}}
\subcaptionbox{\label{flr-tube}}{\includegraphics[trim = 5mm 5mm 5mm 5mm,width=.49\textwidth]{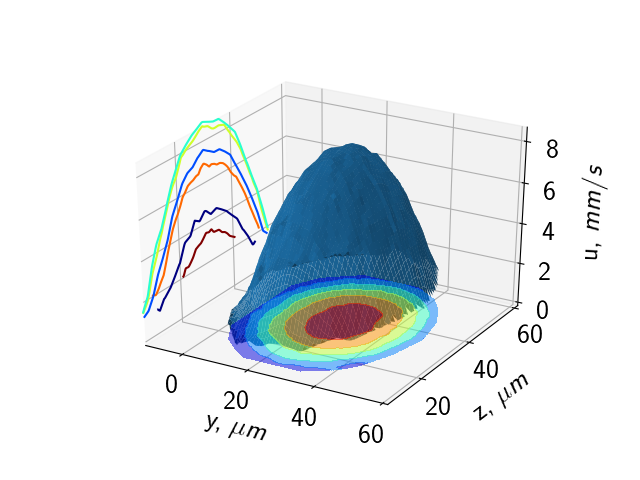}}
\caption{3D plot of velocity profile across a tube cross-section acquired using (a) BF and (b) SPIM-{\textmu}PIV. Selected flow cross-sections plotted on the left of each graph; heatmap of velocities shown in the $yz$ plane of the graph. The analysis was performed using same PIV settings for both datasets, as well as the same optics. The BF results,~\ref{bf-tube}, show flattening of the expected parabolic flow profile in the depth direction, due to depth of correlation effects in the presence of a parabolic flow gradient. In contrast, the SPIM dataset recovers the expected parabolic flow profile as a function of depth, demonstrating the absence of any significant error in the velocity values as a function of depth.}
\label{fig:tube-experiments}
\end{figure}
\subsection{Brightfield vs SPIM-{\textmu}PIV }

For a flow in a tube imaged with volume illumination-{\textmu}PIV systems, DoC effects lead to underestimation of the peak velocity and flow profile broadening \cite{Kloosterman2011}. This is often modelled as having the effect of averaging the velocity over some slab of volume through the sample. To verify the minimal DoC effects in our SPIM-{\textmu}PIV system, we compare it to BF-{\textmu}PIV for imaging flow of 1.04$\si\um$ diameter fluorescent polystyrene beads in a 48$\si\um$ diameter Fluorinated Ethylene Propylene (FEP) tube (refractive index $\approx$ 1.344).

\subsection{Out-of-plane motion effects}
\label{subsec:oopm}
For planar PIV systems, OOPM is one of the potential sources of measurement errors. The effect of particles entering and leaving the thickness of the light sheet reduces the signal peak height and increases background noise in the cross-correlation matrix~\cite[p.176]{Willert2007}, and if steps are not taken to remedy this then the true signal peak eventually becomes indistinguishable from noise. Experimentally, the simplest solution is contract the time difference between laser pulses illuminating the sample between frames, but if this is too short then the frame-to-frame motion (in pixels) becomes unacceptably small.

Other methods discussed in \cite[p.176]{Willert2007} include orienting the sample such that the flow is as much in-plane as possible -- but this is not always feasible in biological samples. Alternatively, for steady flows the statistics of the correlation matrix can be improved by using a technique called \textit{correlation averaging} to combine information from multiple frame-pairs and thus improve tolerance of OOPM.
To investigate how much OOPM the SPIM-{\textmu}PIV can tolerate, we imaged a sample of 1.04$\si\um$ diameter fluorescent beads fixed in 1\% agarose (refractive index $\approx$ 1.343) in a 1.3/1.6$\si\mm$ inner/outer diameter FEP tube, which was scanned through the light sheet to obtain an image stack. We then performed correlation-averaged PIV analysis on image pairs with a known depth difference and examined the accuracy of the results.

\subsection{Blood flow imaging in the zebrafish}
As a proof of concept application, we imaged the flow of fluorescent RBCs (transgenic gata:DsRed line) in the atrium of a 3 day old zebrafish heart. The fish was anaethesized using $\approx$170$\si{\mg\per\l}$
Tricaine Methanesulfonate. Simultaneous to the fluorescence imaging, the heart was imaged in the BF modality (as shown in Fig.~\ref{fig:spim_diagram}) to obtain phase information allowing us to assign heartbeat phases to each PIV frame pair. This then enabled us to use correlation averaging techniques to combine partial flow information from multiple successive heartbeats in a statistically rigorous manner in order to reduce the effects of noise and RBC sparsity on our measurements. The fish was optically sectioned every 8$\si\um$, which for a basic PIV analysis using $\sim 16\si\um$ interrogation windows actually yields a flow field with a $z$ resolution that is double the $xy$ lateral resolution for the calculated flow vectors.

\section{Results and Discussion}
\subsection{Validation experiments}

\paragraph{Flow in a circular tube}
For a nominal flow of 0.5$\si{\ul\per\min}$, the BF-{\textmu}PIV results estimated a peak flow value that was $\approx$9\% lower than the SPIM-{\textmu}PIV system. However, the most important difference is the shape of the recovered flow profile, as shown in Fig.~\ref{fig:tube-experiments}, which is significantly flattened, underestimates the peak velocity, and fails to reproduce the no-slip boundary condition at the tube walls.
\begin{figure}[b]
	\centering
	\includegraphics[width =\textwidth]{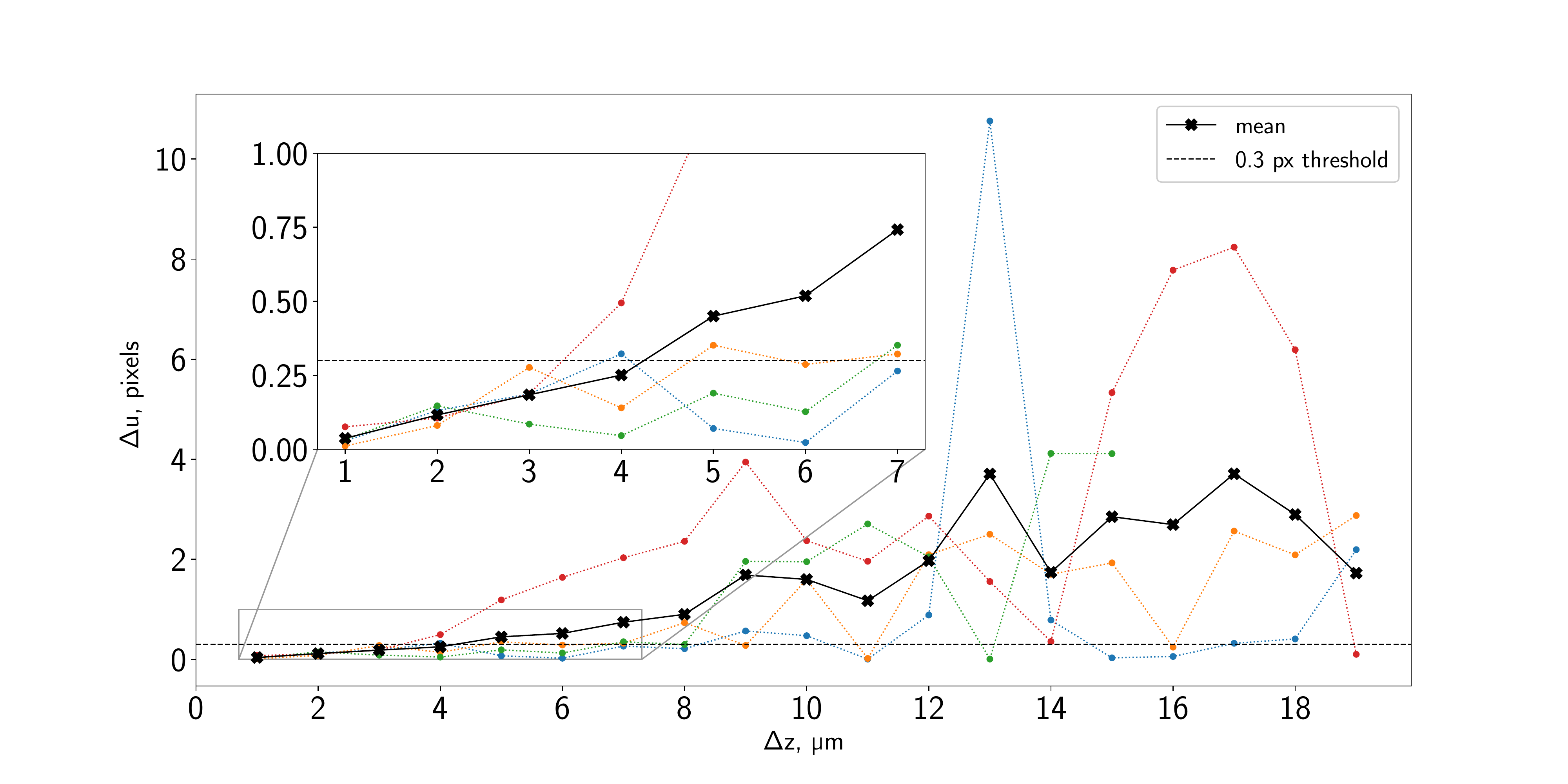}
    \caption{Velocity measurement error due to out-of-plane motion (OOPM). By correlation averaging 4 different datasets (finely dashed lines with dots) with different depth motion, SPIM-{\textmu}PIV data can remain reliable up to 4$\si\um$ shift. In principle, correlating for longer should extend this robust range.}
     \label{fig:oopm}
\end{figure}
This illustrates that the effect of DOC and the ``weighted depth averaging" due to out-of-plane gradients can be extreme for full volume illumination systems. Accounting for them in the post-processing case is non-trivial
except under very specific and controlled circumstances
\cite{Kloosterman2011},
and imaging complex fluid systems can lead to poor estimation of the real dynamics. In contrast, the SPIM system's DOC is prescribed by the thickness of its light sheet, and any averaging is restricted to a very thin section (a typical light sheet Full Width Half Maximum (FWHM) is $\approx$ 2$\si\um$).
\begin{figure}[b]
	\centering
	\includegraphics[width =.7\textwidth]{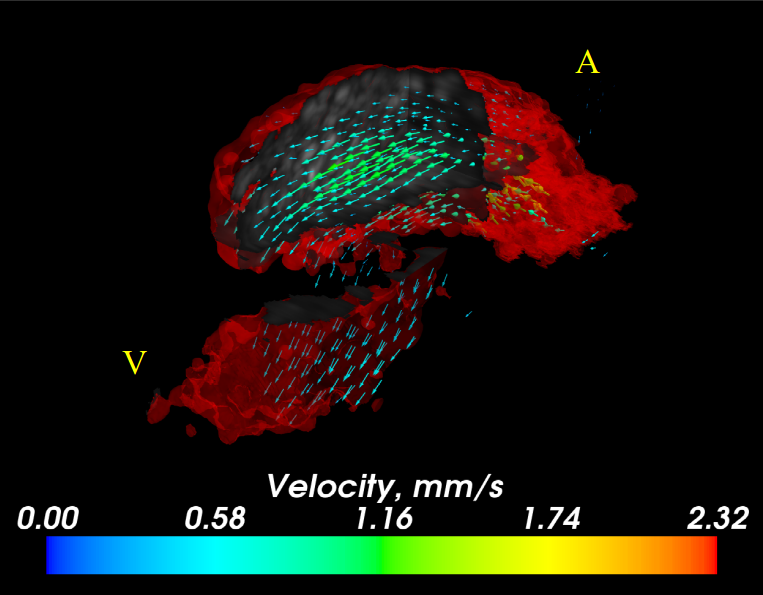}
    \caption{Three dimension-two component (3D-2C) + time blood flow measurement in a 3 day old zebrafish heart. The vectors are shown with square root scaling to better illustrate the dynamic range. In this work, the focus was given to the atrium (indicated by yellow "A") since it is the easier chamber to image at that age. Note that only a small fraction of the ventricle (yellow "V") is captured in this dataset. Figure created with Mayavi \cite{ramachandran2011mayavi}.}
     \label{fig:atrium-cutplane}
\end{figure}
\paragraph{Error due to OOPM}
We show the results of 4 independent correlation averaging results, for different amount of out-of-plane motion, $\Delta z$ in Fig.~\ref{fig:oopm}, where the $y$-axis shows the difference in pixels from the nominal value (the images were synthetically shifted by a fixed value). We chose a 0.3 pixel difference to be a threshold for acceptable error (dashed line in Fig.~\ref{fig:oopm}), a stringent threshold consistent with the engineering PIV literature \cite{Wieneke2015}. With the exception of one dataset (red line in Fig.~\ref{fig:oopm}), the correlation averaged results lie below, or very close to, the threshold up to $\Delta z=4\si\um$ on average, before steeply climbing after 7$\si\um$.

The reason for this tolerance of OOPM when using correlation averaging can be explained in the following way. As described in~\ref{subsec:oopm}, OOPM will dampen the contribution of the ``true" peak, while increasing the number of false peaks (random noise). By correlation averaging, the random noise becomes more evenly distributed, while the signal peak increases, since there are always certain particles common to both frame A and B, up to a certain OOPM value (which all contribute to the true signal peak). The limiting case, above which correlation averaging will not improve the accuracy of measurements, can be visualized as some number of particles just entering the light-sheet in frame A, and those same particles just leaving the light-sheet in frame B. Up to this threshold, increasing the number of frame pairs used in correlation averaging should improve the accuracy.

\subsection{4D blood flow data in the zebrafish atrium}
To resolve the flow over the full heart cycle of the zebrafish, we recorded 27440 PIV frame pairs in total, split over 16 different $z$ planes. After dividing the heartbeat up into phase bins of 0.2 radians, this gave an average of 53 frame pairs per phase bin at each plane, providing ample raw data for correlation averaging to obtain a high quality flow measurement. The results for a phase at mid-filling of the atrium (indicated by letter A) are shown in Fig.~\ref{fig:atrium-cutplane}. By integrating the measured velocity across a plane in the atrium, we found the peak flow rate to be $\approx$2.6$\si{\nano\L\per\s}$.

To consider whether the OOPM effects explained in~\ref{subsec:oopm} might have affected our results, we examined the maximum $y$-component of the velocity throughout the heartbeat.  If we crudely approximate the heart chamber geometry as a radially symmetric contracting tube, we can estimate that the maximum out-of-plane ($z$) velocity (which we cannot directly measure when imaging in a single orientation) should be the same as this maximum $y$ velocity which we \emph{can} measure. We found the $y$-component exceed 1.5\si{\um} OOPM in only 5 of 100 interrogation windows in only one phase bin (confined to one corner of the atrium). We therefore conclude that, in conjunction with our validation results shown in~\ref{fig:oopm}, OOPM effects have not significantly impacted the accuracy of the results.

\section{Conclusion}
We have shown that our SPIM-{\textmu}PIV system represents a powerful platform for {\textmu}PIV measurements since, in contrast to traditional {\textmu}PIV approaches, its light sheet illumination bypasses DOC effects. Furthermore, when used in conjunction with our previously-reported optical gating techniques, we are able to measure high quality depth resolved flow data in the zebrafish heart. We propose that SPIM-{\textmu}PIV is a promising tool for studying biological microfluidic environments, and represents a significant step forward in high-quality quantitative blood flow measurements in small animal models such as the zebrafish.

\section{Acknowledgements}
Zebrafish were kept as stated by the Animals (Scientific Procedures) Act 1986, United Kingdom. We are thankful to Charlotte Buckley, Carl Tucker and Martin Denvir for their help and advice with zebrafish breeding. VZ is supported by a research training support grant from the EPSRC (EP/M506539/1). This work was also partially supported by the British Heart Foundation (NH/14/2/31074), and used equipment funded by the EPSRC (EP/M028135/1) and the Royal Society (RG130249).

\bibliography{report} 
\bibliographystyle{spiebib} 

\end{document}